\documentclass[%
reprint,
superscriptaddress,
aps,
prd,
]{revtex4-2}

\usepackage[toc,page]{appendix}
\usepackage{amsbsy}
\usepackage{amsfonts}
\usepackage{amsopn}
\usepackage{amsmath}
\usepackage{amssymb} 
\usepackage{amstext}
\usepackage{amsthm}
\usepackage{amsxtra}
\usepackage{array} 
\usepackage{booktabs}
\usepackage{bm}
\usepackage{braket} 
\usepackage{cancel}
\usepackage{calc}
\usepackage{comment}
\usepackage{dcolumn}
\usepackage{enumerate}
\usepackage{epsfig}
\usepackage{esint}
\usepackage{extarrows}
\usepackage{footmisc}
\usepackage{graphicx}
\usepackage[colorlinks=true, linkcolor=blue, citecolor=red, urlcolor=magenta]{hyperref}
\usepackage[all]{hypcap}
\usepackage{ifpdf}
\usepackage{indentfirst}
\usepackage{listings}
\usepackage{makeidx}
\usepackage{mathrsfs}
\usepackage{multirow}
\usepackage[displaymath,textmath,sections,graphics,floats]{preview}
\usepackage{pythonhighlight}
\usepackage{simplewick}
\usepackage{slashed}
\usepackage{subfigure}
\usepackage{threeparttable} 
\usepackage{tcolorbox}
\usepackage{url}
\usepackage{xcolor}

\begin{document}

\title{Gravitational waves of GUT phase transition during inflation}

\author{Xi-He Hu}
\email{huxihe23@mails.ucas.ac.cn}
\affiliation{School of Fundamental Physics and Mathematical Sciences, Hangzhou Institute for Advanced Study, UCAS, Hangzhou 310024, China}
\affiliation{Institute of Theoretical Physics, Chinese Academy of Sciences, Beijing 100190, China}
\affiliation{University of Chinese Academy of Sciences, Beijing 100049, China}
\author{Ye-Ling Zhou}
\email{zhouyeling@ucas.ac.cn}
\affiliation{School of Fundamental Physics and Mathematical Sciences, Hangzhou Institute for Advanced Study, UCAS, Hangzhou 310024, China}

\begin{abstract}
  Grand unified theory (GUT) phase transition is generally considered unobservable due to its ultrahigh energy scale, and the monopole problem associated with GUT phase transition is one motivation of inflation. We propose that if a first-order GUT phase transition happens during inflation, the induced gravitational waves (GWs) are redshifted and deformed, and might be observed today in GW observatories.
  We review the formalism of inflated GWs and derive the general deformation function between inflated and uninflated GW spectra in the instant-source or transitory-source application. It is valid for any e-folding number of instant or transitory source. 
  Applying the formalism to GUT phase transition, we find that the e-folding number at 15 or 25 can shift the GWs to 10 Hz or mHz hands, respectively, which might be tested in the future ground-based or space-based interferometers. 
  We further generalise the discussion to inflated GWs via phase transition below the GUT scale. 
  It is worth mentioning that, due to the deformation of the spectrum, the peak of inflated GWs is not simply a redshift of the peak of uninflated GWs.
\end{abstract}

\maketitle

\section{Introduction}

Grand unified theories (GUTs) possess exactly large gauge symmetries that mix the strong and electroweak interactions, and the symmetries become spontaneously broken at an exceedingly high energy scale, the GUT scale. The existence of magnetic monopoles is a universal prediction in GUTs, in spite of the detail of GUT symmetry breaking \cite{Polyakov:1974ek, tHooft:1974kcl}.
Monopoles have masses in general of the same order of the GUT scale. 
They cause cosmological problems if they are generated in the radiation era, since their relic energy density remaining today exceeds the critical energy density of the Universe too much \cite{Kibble:1976sj, Zeldovich:1978wj, Preskill:1979zi}.
Embedding an inflationary era before the radiation era and spontaneously breaking GUT symmetry before or during inflation dilute the monopole number density and thus provide an appealing solution to the GUT monopole problem \cite{Guth:1980zm, Linde:1981mu, Albrecht:1982wi}.  

Gravitational waves (GWs) have been suggested as a complementary probe to GUTs along with proton decay measurements \cite{King:2020hyd}. It utilises the property of $SO(10)$ GUTs that $SO(10)$ contains the gauged $U(1)_{B-L}$ sub-symmetry, and GWs are emitted from cosmic strings, which are generated after the spontaneous breaking of $U(1)_{B-L}$ \cite{Dror:2019syi}. The $U(1)_{B-L}$ symmetry breaking scale is in general split away from the GUT scale, in particular in the non-SUSY model \cite{King:2021gmj}, and connected with the GUT scale via gauge unification \cite{Bertolini:2009qj,Chakrabortty:2019fov,Meloni:2019jcf}. In SUSY GUTs, these two scales might be located nearby with inflation arranged properly \cite{Antusch:2023zjk}, and GWs reduced in the IR band from metastable cosmic strings are expected \cite{Buchmuller:2019gfy,Buchmuller:2020lbh,Buchmuller:2021mbb}, seeing also in \cite{Masoud:2021prr,Fu:2023mdu,Lazarides:2023rqf,King:2023wkm}. 

GWs via cosmological phase transition are motivated in the GUT framework. Specific examples have been carried out in the  
Pati-Salam models \cite{Croon:2018kqn, Huang:2020bbe, Athron:2023aqe} and left-right symmetric models \cite{Brdar:2019fur, Li:2020eun}. To achieve a GW spectrum within the reach of GW observatories such as LIGO/Virgo/KAGRA, a relatively low intermediate scale $\sim 10^{4}$ - $10^{6}$~GeV should be satisfied, which requires appropriate threshold effects in the gauge unification \cite{Croon:2018kqn, Athron:2023aqe}. 
The property of GWs via phase transition during inflation was studied in \cite{An:2020fff}. They pointed out that GWs get redshifted by the inflation, and its power spectrum oscillates with its wave number, making itself distinct from GWs from phase transitions after the inflation. Further developments of the formalism and its applications are found in \cite{An:2022cce, An:2023jxf}. Imprints on the cosmic microwave background via phase transition during inflation was earlier studied in \cite{Jiang:2015qor}.

In the present work, we will consider the GUT phase transition proceeds during the inflation at a certain point with
e-folding number $N_\star$. Different from some well-known inflation models, e.g., \cite{Shafi:1983bd, Lazarides:1995vr}, where the GUT breaking associates with inflationary dynamics, we will assume that the inflation evolves independent from the GUT breaking, and the GUT phase transition is carried out very fast. In such a case, the phase transition will be regarded as an instant or transitory (the duration is non-instant but short) source of GW production, and the inflaton field is simply a background during the phase transition.   

The rest of this paper is organised as follows. In section~\ref{sec:2}, we review the formalism of GWs via phase transition during inflation. A more general formalism, compared with that in \cite{An:2020fff, An:2022cce}, for the deformation function between the inflated GWs and the uninflated GWs (i.e., GWs via phase transition after inflation) is derived. It is valid for any value of the e-folding number $0<N_\star \lesssim 60$. We will further obtain the simplified form of the deformation function in the IR and UV regimes. We apply the formalism to GUT phase transition in section~\ref{sec:3}. We start with the appropriate regime of  $N_\star$ in solving the GUT monopole solution and then concentrate on the property of inflated GWs via GUT phase transition and their testability in GW observatories. Generalisation to phase transition below the GUT scale is given in the end of the section. We conclude in section~\ref{sec:4}.

\section{General formulation of inflated GWs} \label{sec:2}

In this section, we will give a general review on the GWs produced from an instant or transitory source during inflation and their propagation to today. 
We work in the standard picture that the universe begins with an inflation (Inf) period and follows with three eras in the $\Lambda$CDM model, i.e., radiation domination (RD), matter domination (MD) and dark-energy domination ($\Lambda$D) eras. 
We will consider GW genesis via instant source, which happens at a certain time $t_\star$ during the inflation period. 
Then GWs propagate in the inflation + $\Lambda$CDM eras. During its propagation along the expansion of the Universe, the spectrum gets redshifted and deformed. We will show the formalism of the GW spectrum today from  its original spectrum at production. 
Section~\ref{sec:EOMGW} gives the formulation of EOM of GWs in the inflation. The general solution of the GW spectrum from an instant source modified by inflation is given in section~\ref{sec:instant_source}. Its IR and UV behaviours are given in section~\ref{sec:UVIR}. When the duration of sources are non-instant but short, the smearing effects of transitory sources are performed in section \ref{sec:short_source}.

\subsection{Production and propagation of GWs}\label{sec:EOMGW}

The expanding Universe in each era (Inf, RD, MD or $\Lambda$D) can be described by the Friedmann-Lemaitre-Robertson-Walker (FLRW) metric, which in the conformal frame is expressed as
\begin{align}
ds^2 = a^2(\tau)\left[ {\rm d}\tau^2 -{\rm d}{\bf x}^2 \right] \,,
\end{align}
where $a$ is the scale factor. 
We follow the convention $a'={\partial a}/{\partial \tau}$, $\dot{a}={\partial a}/{\partial t}$ with ${\rm d}t = a\, {\rm d}\tau$ understood and it is straightforward to check that the conformal Hubble factor $\mathcal{H}={a'}/{a}$ correlates with the standard Hubble rate $H = {\dot{a}}/{a},$ as $\mathcal{H}=aH$. The conformal Hubble factor ${\cal H}$ in each era  (i.e., $E =$  Inf, RD, MD and $\Lambda$D) of the Universe is given by
\begin{align}
{\cal H}^E = \frac{1}{\tau} \times \{-1,1,2,-1\} \,,
\end{align}
respectively. Note that $\tau$ is convenient to use when studying GW dynamics in a specified era. 
 However, it is not continuously defined when the Universe transits from the end of the last era to the beginning of the next era. 
When we are focusing on the the whole expanding history of the Universe, we can transform  it to the co-moving factor $a$ by default, with $a_0 = a(t_0)$ is the scale factor today.

We regard GWs as small fluctuations in the FLRW background. In the conformal coordinates, the GW metric is given by
\begin{align}
ds^2 = a^2(\tau)\left[ {\rm d}\tau^2 - (\delta_{ij}+h_{ij}(\tau,\mathbf{x})){\rm d}x^i{\rm d}x^j \right]\,,
\end{align}
where we have considered the traceless and transverse gauge for $h_{\mu\nu}$. 
The motion equation of $h_{ij}$ is written to be
\begin{align}
  \begin{aligned}
    &h''_{\mu\nu}(\tau,\mathbf{x})+2\mathcal{H}h'_{\mu\nu}(\tau,\mathbf{x})-\nabla^2h_{\mu\nu}(\tau,\mathbf{x})\\
    &=16\pi G_{\rm N}a^2(\tau) \sigma_{\mu\nu}(\tau,\mathbf{x}) \,,
  \end{aligned} \label{eq:GWEFRW}
\end{align}
where $\sigma_{\mu\nu}$ is the perturbative energy-momentum tensor of sources producing GW in the local Minkowski coordinates. 
Generally, one studies the spectrum of  energy density of GW relying on its momentum.
Therefore, we consider Fourier transforms in conformal coordinates
\begin{align}  
h_{ij}(\tau,\mathbf{x})&=\int\frac{{\rm d}^3 k}{(2\pi)^3} e^{-i\mathbf{k}\cdot\mathbf{x}} \tilde{h}_{ij}(\tau,\mathbf{k}) \,, \nonumber\\
    \sigma_{ij}(\tau,\mathbf{x})&=\int\frac{{\rm d}^3 k}{(2\pi)^3} e^{-i\mathbf{k}\cdot\mathbf{x}} \tilde{\sigma}_{ij}(\tau,\mathbf{k}) \,,    
  \label{eq:FT1}
\end{align}
and then, the equation of motion of GW in the conformal momentum space is 
\begin{align}
  \begin{aligned}
    &\tilde{h}''_{ij}(\tau,\mathbf{k})+2\mathcal{H}\tilde{h}'_{ij}(\tau,\mathbf{k})+k^2\tilde{h}_{ij}(\tau,\mathbf{k})\\
    &=16\pi G_{\rm N}a^2(\tau)\tilde{\sigma}_{ij}(\tau,\mathbf{k})\,.
  \end{aligned}\label{eq:GWEpFRW}
\end{align}

We first ignore the source on the right hand side (RHS) and consider the GW propagation from an preexisting GW metric. The general solution is given by 
\begin{align} \label{eq:h_propagation}
\tilde{h}_{\mu\nu}(\tau,\mathbf{k})=C_{\mu\nu,1} \, h_1(\tau,\mathbf{k}) + C_{\mu\nu,2} \, h_2(\tau,\mathbf{k}),
\end{align}
where $h_1(\tau,\mathbf{k})$ and $h_2(\tau,\mathbf{k})$ are two independent functions of Eq.~\eqref{eq:GWEpFRW} with zero source on the RHS and $C_{\mu\nu,1}$ and $C_{\mu\nu,2}$ are coefficients. 
In a realistic calculation, we have to specify the eras $E$, (for $E =$ Inf, RD, MD, $\Lambda$D), since ${\cal H}$ is different in each era. $h_1(\tau,\mathbf{k})$ and $h_2(\tau,\mathbf{k})$ are given in different forms, as shown in Eq.~\eqref{eq:h_12}.
\begin{widetext}
  \begin{align}
    \begin{aligned}
      &h^{\rm Inf}_1(\tau,\mathbf{k})=\cos k\tau+k\tau\sin k\tau\,, 
      &h^{\rm Inf}_2(\tau,\mathbf{k})&=\sin k\tau-k\tau\cos k\tau\,, \\
      &h^{\rm RD}_1(\tau,\mathbf{k})=\frac{\cos k\tau}{k\tau}\,, 
      &h^{\rm RD}_2(\tau,\mathbf{k})&=\frac{\sin k\tau}{k\tau}\,, \\
      &h^{\rm MD}_1(\tau,\mathbf{k})=\frac{\cos k\tau+k\tau \sin k\tau}{(k\tau)^3}\,, 
      &h^{\rm MD}_2(\tau,\mathbf{k})&=\frac{\sin k\tau-k\tau \cos k\tau}{(k\tau)^3}\,, \\
      &h^{\Lambda{\rm D}}_1(\tau,\mathbf{k})=\cos k\tau+k\tau\sin k\tau\,,
      &h^{\Lambda{\rm D}}_2(\tau,\mathbf{k})&=\sin k\tau-k\tau\cos k\tau\,.
    \end{aligned}\label{eq:h_12}
  \end{align}
\end{widetext}
The relevant parameters $C^E_{\mu\nu,1}$ and $C^E_{\mu\nu,2}$ are not fully free. Matching from the end of a previous era to the beginning of the next era should be taken into account. For example, $h^{\rm Inf}_{ij}$ at the end of inflation and $h^{\rm RD}_{ij}$ at the beginning of RD follows the matching conditions,
\begin{align}
  \begin{aligned}
    \tilde{h}_{ij}^{\rm Inf}(\tau^{\rm Inf},\mathbf{k}) \big|_{\rm Rh}&=\tilde{h}_{ij}^{\rm RD}(\tau^{\rm RD},\mathbf{k}) \big|_{\rm Rh}\,, \\
    \partial_t \tilde{h}_{ij}^{\rm Inf}(\tau^{\rm Inf},\mathbf{k})\big|_{\rm Rh} &= \partial_t \tilde{h}_{ij}^{\rm RD}(\tau^{\rm RD},\mathbf{k})\big|_{\rm Rh}\,. \label{eq:matching}
  \end{aligned}
\end{align}
leading to the correlation among $C^E_{\mu\nu,1}$ and $C^E_{\mu\nu,2}$ between two adjacent eras. Note that the differential with respect to $t$ in the above equation can be replaced by the differential with respect to $\tau$ if the variation of the scale factor in the reheating period is ignored. We remind that $\tau$ does not vary continuously during the transition from the previous era to the next era, e.g., from the end of Inf to the beginning of RD. Thus, we have used two different variables $\tau^{\rm Inf}$ and $\tau^{\rm RD}$ to indicate two discontinuity of conformal time in different eras.  And the matching condition here has ignored the potential modification induced by some effects during the transition.

We then solve Eq.~\eqref{eq:GWEpFRW} in the presence of a source on the RHS. The equation has a general solution expressed by the Green function, 
\begin{align} 
  \begin{aligned}
    \tilde{h}_{ij}(\tau,\mathbf{k})= 16\pi G_{\rm N} \int&{\rm d}\tau'\,\theta(\tau-\tau')\, a^2(\tau')\, \tilde{\sigma}_{ij}(\tau',\mathbf{k})\\
     &\times \mathcal{G}(\tau,\tau';\mathbf{k})\,,
  \end{aligned}\label{eq:h_production}
  \end{align}
where 
\begin{align}
  \begin{aligned}
    \mathcal{G}(\tau,\tau';\mathbf{k})=&\left[\frac{\partial h_1}{\partial\tau}-\frac{h_1}{h_2}\frac{\partial h_2}{\partial\tau}\right]^{-1}_{\tau=\tau'} h_1(\tau,\mathbf{k})\\
    & + \left[\frac{\partial h_2}{\partial\tau}-\frac{h_2}{h_1}\frac{\partial h_1}{\partial\tau}\right]^{-1}_{\tau=\tau'} h_2(\tau,\mathbf{k})
  \end{aligned}\label{eq:GF_general}
\end{align}
and $h_1$ and $h_2$ have been given in Eq.~\eqref{eq:h_12}. Here, the Green function $\mathcal{G}(\tau,\tau';\mathbf{k})$ depends not just on the time difference $\tau-\tau'$ but on both $\tau$ and $\tau'$ individually since the time-translation invariance does not hold in the expanding universe.

In this work, we will consider the GW is produced during the inflation, where Eq.~\eqref{eq:h_production} will be used. 
Once the GW is produced,  its propagation in the expanding Universe follows the equation Eq.~\eqref{eq:h_propagation}, with matching conditions between different eras, such as Eq.~\eqref{eq:matching}, to be considered. 

Regarding GWs as a stochastic background, the energy density is given by 
\begin{align}
  \begin{aligned}
    \rho_{\rm GW}&= \frac{1}{32\pi G_{\rm N}a^2(t)}\Braket{\left|h'_{ij}(\tau,\mathbf{x})\right|^2}\\
    &=\frac{1}{32\pi G_{\rm N}a^2(t)} \int_{T_\tau} \frac{{\rm d}\tau}{T_\tau} \int \frac{{\rm d}^3 \mathbf{k}}{(2\pi)^3 V} \left|h'_{ij}(\tau,\mathbf{k})\right|^2 \,,
  \end{aligned}
\end{align}
where $V$ is the conformal volume. Here, the average along the relevant  conformal period $T_\tau$ for GW genesis is considered. It applies in the condition that the oscillation duration is much smaller than ${\cal H}$.
Thus, the energy density per logarithmic momentum in conformal coordinates is 
\begin{align}  
\frac{{\rm d} \rho_{\rm GW}}{{\rm d} \log k}= \frac{1}{64\pi^3G_{\rm N}}\frac{k^3}{V}\frac{1}{a^2(t)} \int_{T_\tau}\frac{{\rm d}\tau}{T_\tau}\,\left|\tilde{h}'_{ij}(\tau,\mathbf{k})\right|^2\,,  
  \label{eq:rhoGW1}
\end{align}
where $T_\tau$ is duration of GW genesis in the conformal unit.
The spectrum of the GW background in terms of the physical frequency $f$ in the current epoch $t_0$ is defined as
\begin{align}
h^2 \Omega_{\rm GW}(f)=\frac{h^2}{\rho_{\rm c}} \frac{{\rm d} \rho_{\rm GW}}{{\rm d} \log k}\Big|_{t=t_0,k= 2\pi a_0 f}\,,
\label{eq:Omega1}
\end{align}
where $\rho_{\rm c} = 3H_0^2/(8\pi G_{\rm N})$ is critical energy density today.

\subsection{Inflated GW spectrum in the instant-source approximation} \label{sec:instant_source}

In this subsection, we approximate the GW source as an instant source and calculate the resulting GW spectrum. This approach applies to the case that the GW production is much faster than the Hubble expansion. 

We fix the instant source at time $\tau_\star$ in the inflationary era, i.e, 
\begin{align}
  \tilde{\sigma}_{\mu\nu}(\tau,\mathbf{k})=\frac{\delta(\tau-\tau_\star)}{a(\tau_\star)}\, \tilde{\sigma}_{\mu\nu}(\mathbf{k}) \,.
\end{align}
It is convenient to obtain the Green function in Eq.~\eqref{eq:GF_general} (with $E=$ Inf) from time $\tau_\star$ to $\tau$ ($\tau > \tau_\star$) as
\begin{align} 
  \begin{aligned}
    \mathcal{G}_{\rm Inf}(\tau,\tau_\star;\mathbf{k})=& \left[\frac{1}{k^2\tau_\star^2}+\frac{a(\tau_\star)}{a(\tau)}\right] \frac{\sin[k(\tau-\tau_\star)]}{k}\\ &+\left[1-\frac{a(\tau_\star)}{a(\tau)}\right]\frac{\cos[k(\tau-\tau_\star)]}{k^2\tau_\star} \,,
  \end{aligned}
\end{align}
where $\tau_\star$ can be expressed in terms of $a_\star$ as $\tau_\star=-(H_{\star}a_\star)^{-1}$. Here $H_\star$ is the Hubble rate at the time of instant source, and since source is during inflation, we have identified $H_\star$ as the constant Hubble expansion rate during inflation.
The GW metric, induced by the instant source,
in the later period of inflation is expressed as
\begin{align} 
  \begin{aligned}
    \tilde{h}_{ij}^{\rm Inf}(\tau,\mathbf{k})&= 16\pi G_{\rm N}a_\star\, \tilde{\sigma}_{ij}(\mathbf{k})\, \mathcal{G}_{\rm Inf}(\tau,\tau_\star;\mathbf{k})\\
    &= 16\pi G_{\rm N}\tilde{\sigma}_{ij}(\mathbf{k}) \,h_{\rm Inf}(\tau,\mathbf{k}),\\ 
    &(\tau^{\rm Inf}_{\rm i}<\tau_\star<\tau<\tau^{\rm Inf}_{\rm f}) 
  \end{aligned}\label{eq:hij_inf}
\end{align}
where $\tau^{\rm Inf}_{\rm i}$ and $\tau^{\rm Inf}_{\rm f}$ are initial and final values of the conformal time in the inflation era, respectively, and $h_{\rm Inf}(\tau,\mathbf{k})$ is a scalar function defined as 

\begin{align}
  \begin{aligned}
    h_{\rm Inf}(\tau,\mathbf{k}) =& \frac{a_\star}{k} \left\{ \frac{a_\star H_{\star}}{k}\left(\frac{a_\star}{a(\tau)}-1\right)\cos[k(\tau-\tau_\star)] \right.\\
     &\quad\,  \left. + \left(\frac{a_\star^2H_{\star}^2}{k^2}+\frac{a_\star}{a(\tau)}\right)\sin[k(\tau-\tau_\star)] \right\}. 
  \end{aligned}
 \label{eq:h_inf}
\end{align}

Then, we match the GW metric from the final time of inflation $\tau^{\rm Inf}_{\rm f}$ to the initial time of RD $\tau^{\rm RD}_{\rm i}$. Using $t(\tau^{\rm Inf}_{\rm f})=t(\tau^{\rm RD}_{\rm i})=t_{\rm Rh}$, 
the matching condition in Eq.~\eqref{eq:matching} becomes 
\begin{align}
  \begin{aligned}
    \tilde{h}^{\rm Inf}_{ij} \big|_{\tau^{\rm Inf}_{\rm f}}  &= C^{\rm RD}_{ij,1}\,h^{\rm RD}_1 \big|_{\tau^{\rm RD}_{\rm i}} + C^{\rm RD}_{ij,2}\,h^{\rm RD}_2 \big|_{\tau^{\rm RD}_{\rm i}} \,, \\
    \partial_\tau\tilde{h}_{ij}^{\rm Inf} \big|_{\tau^{\rm Inf}_{\rm f}} &= C^{\rm RD}_{ij,1}\, \partial_\tau \tilde{h}_1^{\rm RD} \big|_{\tau^{\rm RD}_{\rm i}} + C^{\rm RD}_{ij,2}\, \partial_\tau \tilde{h}_2^{\rm RD} \big|_{\tau^{\rm RD}_{\rm i}} \,.
  \end{aligned}\label{eq:match4}
\end{align}
$C^{\rm RD}_{ij,1}$ and $C^{\rm RD}_{ij,2}$ are solved to be
\begin{widetext}
\begin{align}
  \begin{aligned}
    C^{\rm RD}_{ij,1}&=16\pi G_{\rm N}\tilde{\sigma}_{ij}(\mathbf{k})\frac{h_{\rm Inf}(\frac{-1}{H_{\star}a_{\rm Rh}},\mathbf{k})\times(h^{\rm RD}_2)'(\frac{2t_{\rm Rh}}{a_{\rm Rh}},\mathbf{k}) - h'_{\rm Inf}(\frac{-1}{H_{\star}a_{\rm Rh}},\mathbf{k})\times h^{\rm RD}_2(\frac{2t_{\rm Rh}}{a_{\rm Rh}},\mathbf{k})}{h^{\rm RD}_1(\frac{2t_{\rm Rh}}{a_{\rm Rh}},\mathbf{k})\times(h^{\rm RD}_2)'(\frac{2t_{\rm Rh}}{a_{\rm Rh}},\mathbf{k}) - (h^{\rm RD}_1)'(\frac{2t_{\rm Rh}}{a_{\rm Rh}},\mathbf{k})\times h^{\rm RD}_2(\frac{2t_{\rm Rh}}{a_{\rm Rh}},\mathbf{k})}\,, \\
    C^{\rm RD}_{ij,2}&=16\pi G_{\rm N}\tilde{\sigma}_{ij}(\mathbf{k})\frac{h_{\rm Inf}(\frac{-1}{H_{\star}a_{\rm Rh}},\mathbf{k})\times(h^{\rm RD}_1)'(\frac{2t_{\rm Rh}}{a_{\rm Rh}},\mathbf{k}) - h'_{\rm Inf}(\frac{-1}{H_{\star}a_{\rm Rh}},\mathbf{k})\times h^{\rm RD}_1(\frac{2t_{\rm Rh}}{a_{\rm Rh}},\mathbf{k})}{h^{\rm RD}_2(\frac{2t_{\rm Rh}}{a_{\rm Rh}},\mathbf{k})\times(h^{\rm RD}_1)'(\frac{2t_{\rm Rh}}{a_{\rm Rh}},\mathbf{k}) - (h^{\rm RD}_2)'(\frac{2t_{\rm Rh}}{a_{\rm Rh}},\mathbf{k})\times h^{\rm RD}_1(\frac{2t_{\rm Rh}}{a_{\rm Rh}},\mathbf{k})}\,.
  \end{aligned}
\end{align}
\end{widetext}
Therefore, the GW metric during RD period is converted to
\begin{align}
  \tilde{h}^{\rm RD}_{ij}(\tau,\mathbf{k})=16\pi G_{\rm N}\tilde{\sigma}_{ij}(\mathbf{k})\times h_{\rm RD}(\tau,\mathbf{k})\,. \quad  (\tau^{\rm RD}_{\rm i}<\tau) \label{eq:hRD1}
\end{align}
Here, $h_{\rm RD}(\tau,\mathbf{k})$ is a scale function defined as
\begin{gather}
  h_{\rm RD}(\tau,\mathbf{k})=h_0(\tau,\mathbf{k})+h_1(\tau,\mathbf{k}) \,, \label{eq:h_RD}
\end{gather}
where $h_0(\tau,\mathbf{k})$ and $h_1(\tau,\mathbf{k})$ are
\begin{align}
  &\begin{aligned}
    h_0(\tau,\mathbf{k})=\ & \frac{-a_{\rm Rh}^2}{a(\tau)}\frac{1}{a_\star H_{\star}} \times \frac{\sin[k\tau - y \epsilon]}{y^3} \\
    &\ \times \left\{ \cos[y(1\!-\!\epsilon)] - \frac{\sin[y(1\!-\!\epsilon)]}{y} \right\} \,,
  \end{aligned}\nonumber
  \\
  &\begin{aligned}
    h_1(\tau,\mathbf{k})=\ &\frac{-a_{\rm Rh}^2}{a(\tau)}\frac{y \epsilon}{a_\star H_{\star}} \times \left\{ \frac{1-\epsilon}{y^3} \cos[k\tau +y(1\!-\!2\epsilon)] \right.\\
    & \quad \left. - \left(\frac{1+y \epsilon}{y^4} \right) \sin[k\tau +y(1\!-\!2\epsilon)] \right\} \,,
  \end{aligned}\nonumber 
\end{align}
  with  
\begin{align}
  y = \frac{k}{a_\star H_{\star}} \,, \quad \epsilon = \frac{a_\star}{a_{\rm Rh}} \,. \nonumber
\end{align}
Here we have specified $h_{\rm RD}$ into two terms. The first term $h_0(\tau,\mathbf{k})$ dominates $h_{\rm RD}(\tau,\mathbf{k})$ for $y\epsilon\ll 1$, and the second term $h_1(\tau,\mathbf{k})$ dominates $h_{\rm RD}(\tau,\mathbf{k})$ for $y\epsilon\gg 1$. 

The GW energy density is obtained from the derivative 
\begin{align}
  \partial_\tau \tilde{h}^{\rm RD}_{ij}(\tau,\mathbf{k}) = 16\pi G_{\rm N}\tilde{\sigma}_{ij}(\mathbf{k})\times \partial_\tau h_{\rm RD}(\tau,\mathbf{k}) \,.
\end{align}
From the above, it is straightforward to obtain the GW energy density per $\log k$ as
\begin{align}  
  \frac{{\rm d}\rho_{\rm GW}}{{\rm d}\log k} &= \frac{{\rm d}\rho_{\rm GW}^{\rm flat}}{{\rm d}\log k} \times \frac{a_{\rm Rh}^4}{a^4(t)} \times {\cal S}(t,k)\,, \quad (t^{\rm Inf}_{\rm f}< t\le t_0)
  \label{eq:rhoGW2}
\end{align}
where
\begin{align} \label{eq:rho_flat}
\frac{{\rm d}\rho_{\rm GW}^{\rm flat}}{{\rm d}\log k} = \frac{2G_{\rm N}}{\pi}\frac{k^3}{V} \left|\tilde{\sigma}_{ij}(\mathbf{k})\right|^2
\end{align}
is the GW energy density per $\log k$ just in flat spacetime. This is obtained by setting ${\cal H} = 0$ in Eq.~\eqref{eq:GWEpFRW}. Then the metric is solved to be $h_{ij}^{\rm flat} (\tau, \mathbf{k}) = 16\pi G_{\rm N} \tilde{\sigma}_{ij}(\mathbf{k}) {\sin[k(\tau - \tau_\star)]}/{k}$, and Eq.\eqref{eq:rho_flat} is obtained just following the definition in Eq.~\eqref{eq:rhoGW1}. The superscript ``flat'' represents the GW spectrum obtained in the flat spacetime.
And,
\begin{widetext}
  \begin{align}
  \begin{aligned}
    \mathcal{S}(t, k)=&\left(1+\frac{a_{\rm Rh}^4}{a^2(t) a_\star^2y^2}\right) \left\{ \left[ \frac{\cos[y(1-\epsilon)]}{y^2} -  \frac{\sin[y(1-\epsilon)]}{y^3} \right]^2 \right.\\
            &\ \left. +\,  \epsilon \left[\frac{1}{y^2}+{2\epsilon-1}\right] \frac{\sin[2y(1-\epsilon)]}{y^3} - \epsilon \left[\frac{2-\epsilon}{y^2}+{\epsilon}\right] \frac{\cos[2y(1-\epsilon)]}{y^2}   + \epsilon^4 \left(\frac{1}{y^2}+1\right)  \right\}\,,
  \end{aligned}\label{eq:St}
\end{align}
\end{widetext}
is a momentum-dependent deformation function modifying the shape of the GW spectrum via inflation. 
The second term $\frac{a_{\rm Rh}^4}{a^2(t) a_\star^2y^2}$ in the parentheses arises from the derivative of the co-moving factor $a(t)$ of Eq~\eqref{eq:h_RD}. By fixing the time at today ($a(t_0)=a_0$), $\frac{a_{\rm Rh}^2}{a_0 a_\star y}\ll 1$ is in general satisfied for frequency $f\ge 10^{-18}$~Hz.
In the above expression, the oscillation in the metric (via the $\sin(k\tau)$ and $\cos(k\tau)$ terms) has been averaged following the integration $\int_{T_\tau} d \tau$. 
On the other hand, as EOMs of GWs in different era are different, matching from the end of previous period to the beginning of next period, mainly the end of inflation to the beginning of RD, leads to oscillation in the energy spectrum of GW background. 

The next is to consider GWs propagating from RD to MD, and then that from MD to $\Lambda$D eras. We remind that $\tau$ is also discontinuous during the transition from RD to MD, as well as MD to $\Lambda$D. Thus, matching between different eras, similar to Eq.~\eqref{eq:matching}, must be accounted. 
For GW frequency of our interests, i.e. $f\gtrsim 10^{-9}$~Hz, as the GWs are always inside the horizon during RD, MD and $\Lambda$D, the matching is trivial and we dismiss it in this study.
On the other hand, for lower frequency, in particular, for $f\le 10^{-14}$~Hz, the matching at the matter-radiation equality should be considered. 

\begin{figure*}[t!]
  \centering
      \includegraphics[width=.9\textwidth]{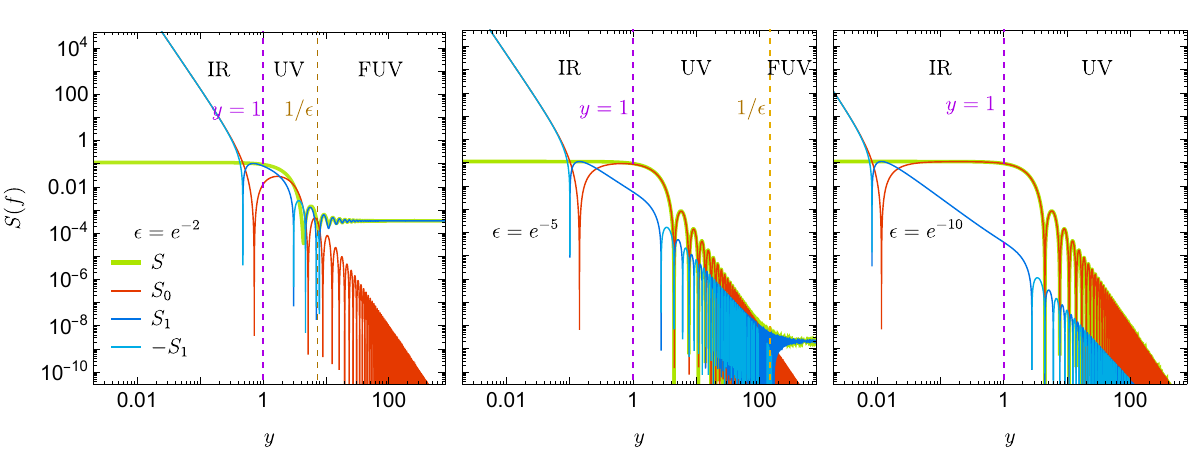}
      \caption{Deformation function along the frequency with inputs $T_{\rm Rh}=10^{15}$~GeV, $\epsilon=e^{-2}$ (left), $e^{-5}$ (middle) and $e^{-10}$ (right) in the instant source approximation. The IR, UV and FUV regions are defined in Eq.~\eqref{eq:IR_UV_FUV}.}\label{fig:deformation}
\end{figure*}

We further compare the presently inflated GW spectrum $\Omega_{\rm GW}(f)$ with the uninflated spectrum $\widetilde{\Omega}_{\rm GW}(\tilde{f})$. The latter, more precisely, is the GW spectrum generated from the same instant source,  in the radiation era after the reheating. We fix source in the beginning of the RD era, just after the reheating. The hypothetical frequency not redshifted by inflation, $\tilde{f}$, correlates with the physical frequency $f$ via $\tilde{f}=f\,{a_{\rm Rh}}/{a_\star}$. 
The correlation between inflated and uninflated GW spectrums is straightforwardly obtained as
\begin{align}  
   h^2 \Omega_{\rm GW}(f)&=h^2 \widetilde{\Omega}_{\rm GW}(\tilde{f})\Big|_{\tilde{f}=f\frac{a_{\rm Rh}}{a_\star}} \times S(f)\,.
      \label{eq:hsqOmega_master}
\end{align}
Here, 
\begin{align}
   h^2 \widetilde{\Omega}_{\rm GW}(\tilde{f})=\frac{h^2}{\rho_c} \frac{{\rm d} \rho_{\rm GW}^{\rm flat}}{{\rm d} \log k} \times \frac{a_{\rm Rh}^4}{a_0^4} \label{eq:hsqOmega_tilde}
\end{align}
is the uninflated GW spectrum which is supposed to be observed today. 
As we assumed instant source happens in the beginning of the radiation era, thus the scale factor have been replaced by $a_{\rm Rh}$. 
$S(f)$ is the momentum-dependent deformation function with time set at today, $S(f) \equiv {\cal S}(t_0, k\!=\!2\pi a_0 f)$. We further divided it into two terms  
\begin{align}
  S(f)&=S_0(f) + S_1(f) \, \label{eq:S}
\end{align}
with
\begin{align}
  &S_0(f)= \left\{ \frac{\cos[y(1-\epsilon)]}{y^2} -  \frac{\sin[y(1-\epsilon)]}{y^3} \right\}^2  \,, \label{eq:S0} \\
  &\begin{aligned}
    S_1(f)=&  y \epsilon \times \left\{ \left[\frac{1}{y^2}+{2\epsilon}-1\right] \frac{\sin[2y(1-\epsilon)]}{y^4} \right.\\
    &\left. - \left[\frac{2-\epsilon}{y^2}+{\epsilon}\right] \frac{\cos[2y(1-\epsilon)]}{y^3} + \frac{\epsilon^3}{y} \left(\frac{1}{y^2}+1\right)  \right\} \,,
  \end{aligned}\label{eq:S1}
\end{align}
where 
\begin{align}
  y=\frac{2\pi a_0 f}{a_\star H_{\star}} \,.
\end{align}
$S_0(f)$ is the leading contribution for $y \epsilon \ll 1$, which is directly derived via $h_0(\tau, \mathbf{k})$ in Eq.~\eqref{eq:h_RD}. This is exactly the result listed in \cite{An:2020fff}.
$S_1(f)$ includes the rest terms. In the high frequency band $y \epsilon \gg 1$, $S_1(f)$ can be dominant and thus, we keep it in our formula. 
We show in Fig.~\ref{fig:deformation} the comparison of $S_0(f)$ and $S_1(f)$ along the frequency. More discussions are given in the next subsection. 

\subsection{IR and UV behaviour of the inflated GW spectrum}\label{sec:UVIR}

We discuss in more details on IR and UV behaviours of GW spectrum obtained in the last subsection. The IR and UV bands of the inflated GW spectrum are distinguished by comparing the physical GW wavelength $\lambda_\star = 2 \pi a_\star / k$  at the time of GW produced with the Hubble distance $H_{\star}^{-1}$. $\lambda_\star \gg H_{\star}^{-1}$ refers to the IR band and on the contrary is the UV band. In addition, we specify the Far UV (FUV) band if the physical wavelength at the end of the inflation, which has been redshifted by a factor $a_{\rm Rh} / a_\star$, is much shorter than the Hubble distance at that time, i.e., $\lambda_\star a_{\rm Rh} / a_\star \ll H_{\star}^{-1}$. 
The three regimes expressed in frequency today are respectively given by
\begin{align}
  \begin{aligned}
    \text{IR}:& &   k \ll& a_\star H_{\star} &                                                       &\Rightarrow                &  f \ll& \frac{a_\star H_{\star}}{2\pi a_0} \,, \\
    \text{UV}:& &   \hspace{-5pt}a_\star H_{\star} \ll k& \ll a_{\rm Rh} H_{\star}\hspace{-7pt} &   &\Rightarrow  & \hspace{-3pt}  \frac{a_\star H_{\star}}{2\pi a_0} \ll f& \ll \frac{a_{\rm Rh} H_{\star}}{2\pi a_0} \,, \\
    \text{FUV}:& &  k \gg& a_{\rm Rh} H_{\star} &                                                    &\Rightarrow                &  f \gg& \frac{a_{\rm Rh} H_{\star}}{2\pi a_0}  \,.  
  \end{aligned} \label{eq:IR_UV_FUV}
\end{align}
These regimes are also indicated in Fig.~\ref{fig:deformation}. 
On the other hand, since the parameter $y = \frac{2\pi a_0 f}{a_\star H_{\star}}$ in oscillation terms of the deformation function takes a key role to determine the shape of the spectrum, we can also distinguish the IR, UV, and FUV bands via conditions $y \ll 1$, $1 \ll y \ll 1/\epsilon$ and $y \gg 1/\epsilon$, respectively. 

Given the formulation in the last subsection, we outline the IR, UV and FUV behaviour of the inflated GW spectrum below. 

\begin{itemize}
  \item 
  In the IR band, the wavelength of GWs is outside the horizon during the whole period of inflation and moves back to the horizon after reheating. Therefore, the metric of GWs in the IR band  is hardly diluted by inflation, $\tilde{h}_{ij}^{\rm RD}(\tau,\mathbf{k})|_{y\ll 1}\propto\frac{a_{\rm Rh}^2}{a(t)}$, and from \eqref{eq:h_RD}, we can get
  \begin{align}
    h_{\rm RD}^{\rm IR}\simeq \frac{1}{k}\frac{a_{\rm Rh}^2}{a(t)}\times \frac{1}{3} \sin\left(k\tau-\frac{k}{a_{\rm Rh}H_{\star}}\right) \,.  \label{eq:h_IR}
  \end{align}
  The deformation function of the GW spectrum in the IR band approximates to 
  \begin{align}
    S(f)^{\rm IR} \simeq& \frac{1}{9} \,.  \label{eq:S_IR}
  \end{align}
  The energy density of this case is not strongly diluted and is almost the same as it without inflation except the suppression factor $1/9$. In the above approximation, we have assumed $\epsilon \ll 1$ which is in general true during the inflation. If it does not hold, the factors $1/3$ in Eq.~\eqref{eq:h_IR} and $1/9$ in Eq.~\eqref{eq:S_IR} should be replaced by $(1+2\epsilon^2)/3$ and $(1+2\epsilon^2)^2/9$, respectively. 
  \item
  In the UV but not the FUV band, the wavelength is inside the horizon during the GW production, and then get redshifted outside the horizon before the end of the inflation. It moves back to the horizon at later time after the inflation.
  The GW metric approximates to 
  \begin{align}
    \hspace{25pt} h_{\rm RD}^{\rm UV}\simeq \frac{-1}{k}\frac{a_{\rm Rh}^2}{a(\tau)} \frac{\cos[y(1-\epsilon)]}{y^2} \sin(k\tau- y \epsilon) \,. \label{eq:h_UV}
  \end{align}
  Again we have assumed $\epsilon \ll 1$. The spectrum deformation function approximates to 
  \begin{align}
    S(f)^{\rm UV} \simeq& \frac{\cos[2y(1-\epsilon)] + 1}{2y^4} \,.  \label{eq:S_UV}
  \end{align}
  As we have not moved to the FUV regime, $y \epsilon < 1$, $S_0(f)$ still dominates the deformation function $S(f)$. Thus, the oscillation behaviour is important in this regime.
  \item
  In the FUV band, the wavelength of GWs is always inside the horizon. Hence, the GW metric gives 
  \begin{align}
    \begin{aligned}
      h_{\rm RD}^{\rm FUV}\simeq \frac{1}{k}\frac{a_\star^2}{a(t)} \sin[k\tau + y (1-2\epsilon)] \,.
    \end{aligned}
  \end{align}
  The deformation function provides only a suppression factor, 
  \begin{align}
    S(f)^{\rm FUV} \simeq& \frac{a_\star^4}{a_{\rm Rh}^4} \,.  \label{eq:S_FUV}
  \end{align}     
  Due to the quartic dependence, a small fraction $a_\star /a_{\rm Rh}$ leads to a large suppression to the FUV regime. Once the GW radiation happens significantly before the end of the inflation, the GWs in the FUV band will be highly diluted and thus is almost not observable.  
\end{itemize}
The energy density suppressed by $a_{\rm Rh}^{-4}$ in the FUV band is naturally understood as in this regime, GWs appear as radiations. However, this is not true in other regimes where GWs have longer wavelength outside the horizon at least in a certain period of the inflation. 
It is highlighted that the discussion in the above depends only on the instant-source approximation. On the other hand, if the instant-source assumption works in a realistic dynamics, these discussions apply, in spite of the explicit form of GW source. In the coming section, we will focus on GUT phase transition and show more features of the spectrum of its inflated GWs.

\subsection{Smearing effect induced by non-instant but transitory source} \label{sec:short_source}

\begin{figure*}[t!]
  \centering
      \includegraphics[width=.9\textwidth]{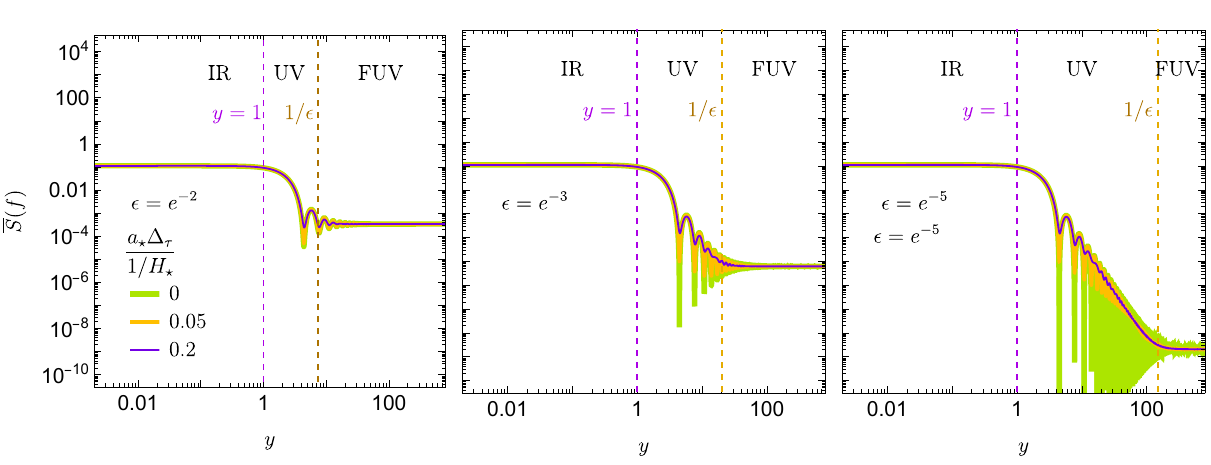}
      \caption{Smeared deformation function with inputs  $\epsilon=e^{-2}$ (left), $e^{-3}$ (middle) and $e^{-5}$ (right) from transitory sources. The normalised time duration $\frac{a_\star\Delta_\tau}{1/H_\star}=0$, 0.05, 0.2 are compared. In particular, $\frac{a_\star\Delta_\tau}{1/H_\star}=0$ refers to the instant source and $\overline{S}(f)$ is same as $S(f)$.}\label{fig:smeared_deformation}
\end{figure*}  

If the duration of sources of GW is same magnitude of cosmic expansion, the sources are non-instant, so the energy density of GW is not simply expressed as Eq~\eqref{eq:hsqOmega_master}.
However, for non-instant but transitory sources of GW, Eq~\eqref{eq:hsqOmega_master} is appropriate.
In this case, deformation function is smeared and written as $\overline{S}(f)$
\begin{align}
  \overline{S}(f)=\frac{1}{\Delta_y}\int_{\bar{y}-\Delta_y/2}^{\bar{y}+\Delta_y/2}{\rm d}y\, S(f) \Bigg|_{\bar{y}=\frac{2\pi a_0 f}{a_\star H_\star}}\, , \label{eq:Sbar}
\end{align}
where $\Delta_y=\frac{a_\star\Delta_\tau}{1/H_\star}\bar{y}$ and $\Delta_\tau$ is conformal duration of sources of GW.
Actually, the analytic expression of $\overline{S}(f)$ is complex but exisits.
Hence, we show the value of $\overline{S}(f)$ with different duration of sources of GW on Fig.~\ref{fig:smeared_deformation}.
The oscillation of smeared deformation function reduces rapidly, which is due to the average of $S(f)$ about $y$.
From Fig.~\ref{fig:deformation} and section~\ref{sec:UVIR}, we can simply discuss $\overline{S}(f)$ in three bands,
\begin{align}
  \overline{S}(f)\simeq\begin{cases}
    S(f) & y\ll 1 \text{ or } y\gg 1/\epsilon\\
    \overline{S}_0(f) & 1\ll y\ll 1/\epsilon \\ 
  \end{cases}\, ,
\end{align}
where $\overline{S}_0(f)$ is approximated as
  \begin{align}
    \begin{aligned}
      \overline{S}_0(f)\simeq&\, \frac{1}{\bar{y}^4\Delta_y}\int_{\bar{y}-\Delta_y/2}^{\bar{y}+\Delta_y/2}\! {\rm d}y \left\{\cos[y(1-\epsilon)] \!-\!  \frac{\sin[y(1-\epsilon)]}{y} \right\} \\
      =&\, \frac{1}{\bar{y}^4} \Big\{ \frac{1}{2} +\frac{1}{\Delta_y}\Big[\frac{\sin^2[(\bar{y}-\Delta_y/2)(1-\epsilon)]}{\bar{y}-\Delta_y/2} \nonumber\\
      &-\frac{\sin^2[(\bar{y}+\Delta_y/2)(1-\epsilon)]}{\bar{y}+\Delta_y/2}\Big] \nonumber\\ 
      &+\frac{\cos[2\bar{y}(1-\epsilon)]\, \sin[\Delta_y(1-\epsilon)]}{2\Delta_y} \Big\} \, .
    \end{aligned}
  \end{align}
If $\epsilon\ll 1$, the above equation is same as \cite{An:2020fff}. The smeared deformation function $\overline{S}(f)$ in the IR, UV and FUV bands can be analytically derived and they are summarised below, 
\begin{align}
    &\bar{S}(f)^{\rm IR} \simeq \frac{1}{9} \,, \nonumber\\
    &\bar{S}(f)^{\rm UV} \simeq \frac{(1-\epsilon) + \cos [2 \bar{y} (1-\epsilon)] \, \sin [\Delta_y(1-\epsilon)]/\Delta_y}{2\bar{y}^4} \,,\nonumber\\
    &\bar{S}(f)^{\rm FUV} \simeq \frac{a_\star^4}{a_{\rm Rh}^4} \,.
\end{align}
$\bar{S}(f)^{\rm IR}$ and $\bar{S}(f)^{\rm FUV}$ are the same as those in the instant approximation in Eqs.~\eqref{eq:S_IR} and \eqref{eq:S_FUV}, respectively. $\bar{S}(f)^{\rm UV}$ matches with $S(f)^{\rm UV}$ in Eq.~\eqref{eq:S_UV} in the limit of zero conformal duration of sources, e.g., $\Delta_y \to 0$.

\section{Application to GUT phase transition}\label{sec:3}

GUTs provide a strong motivation to study phase transition (PT) during the inflation. The monopole problem in GUTs is one important reason to propose an inflationary period before the radiation era. To dilute the energy density of GUT monopoles, the GUT symmetry breaking should take place before or during inflation. Here we will specify the GUT PT is instant or transitory and happens at some point during the inflation, and it is triggered by the varying inflaton field at a certain number of e-folds during the inflation. 
We denote 
\begin{align} \label{eq:e_folds}
N_\star = \log \frac{a_{\rm Rh}}{a_\star} 
\end{align}
as the e-folding number of PT. 
In a certain regime of $N_\star$, we will see that the GW frequency induced by the GUT PT can be redshifted into the observable regime in the foreseeable future and the cosmological monopole problem is solved. 

\subsection{Solution to the monopole problem}

We first give briefly review on how the monopole problem triggered GUT PT is solved by an inflation. It is well-known that a GUT symmetry breaking in the very early Universe, either $SU(5)$ GUT or $SO(10)$ GUT, leads to the generation of monopoles with mass $M_{\rm mono}$ at the GUT scale.  
Given an initial number density of monopoles at their production $n_\star = n_{\rm mono}(t_\star)$, the evolution of monopoles is like the matter, with its number density at later time following
\begin{align}
  n_{\rm mono}(t)=\left(\frac{a(t_\star)}{a(t)}\right)^3 n_\star \,.
\end{align}
The energy density fraction of monopoles today is given by $\Omega_{\rm mono}={M_{\rm mono}\, n_{\rm mono}(t_0)}/{\rho_{\rm c}}$ with $\rho_c = 3H_0^2/(8\pi G_{\rm N})$ the critical energy density today. 
The initial number density is given by the initial correlation length $n_\star = L_{\rm mono,\star}^{-3}$, where $L_{\rm mono, \star}$ is at the same order of the Hubble distance, and here we take $L_{\rm mono, \star} \simeq H_\star^{-1}$ for simplicity. It is well-known that the GUT monopoles cannot be produced in the radiation era. Namely, by taking the monopole mass and the thermal temperature at the GUT scale, we arrive at $\Omega_{\rm mono} \gg 10^{-5}$ which conflicts with the observed Universe. 

Therefore, we consider the PT happening during inflation. Monopoles, which are generated just following the completion of PT,  have the initial correlation length of the same order of the Hubble distance at inflation. Again, we assume $L_{\rm mono,\star} = H_{\star}^{-1}$ for simplicity. Given the e-folding number $N_\star$ in Eq.~\eqref{eq:e_folds}, we obtain $\Omega_{\rm mono}$ is diluted to
\begin{align}
\Omega_{\rm mono}= \frac{8\pi G M_{\rm mono} H_{\star}^3}{3 H_0^2 (1+z_{\rm Rh})^3}  e^{-3 N_\star} \,.
\end{align}
We make naive approximations $H_{\star} \sim H_{\rm Rh} \simeq 0.1 T_{\rm Rh}^2/M_{\rm pl}$ and $M_{\rm mono} \simeq \Lambda_{\rm GUT}/\alpha_{\rm GUT} \sim  10\, T_{\rm Rh}$, and arrive at 
 \begin{align}
    \Omega_{\rm mono}\sim 10^{-5} \left(\frac{T_{\rm Rh}}{10^{15} ~{\rm GeV}}\right)^4  e^{-3(N_\star-15)}  \,. \label{eq:Omega_mono}
 \end{align}
Taking the GUT monopole mass $T_{\rm Rh} \sim 10^{15}$~GeV, we should restrict the e-folding number at the PT $N_\star \gtrsim 15$, such that $\Omega_{\rm mono} \lesssim 10^{-5}$ is small enough to satisfy the cosmological constraints. Note that Eq.~\eqref{eq:Omega_mono} applies also to monopole generated from intermediate symmetry breaking below the GUT scale but sufficiently higher than the electroweak scale. For example, we take the Pati-Salam scale at $10^{14}$~GeV and assume the relevant monopole mass and reheating temperature around the same scale, the requirement $\Omega_{\rm mono} \lesssim 10^{-5}$ relaxes the restriction on the number of e-folds to $N_\star \gtrsim 12$. 

\subsection{Inflated GWs via GUT phase transition}\label{sec:3.2}

A first-order GUT PT can happen in the inflationary era. Different from the widely studied thermal PT in the radiation era, where the temperature plays the crucial role. The slow-roll inflation field may take the role in the PT during inflation.  

Whether a first-order PT can occur depends on the form of the effective potential during the PT. Without knowing the detail of GUT, we do not know the explicit form of the effective potential during the PT. 
Although our latter discussion does not depend on any details of the form of the potential for instant or transitory PT, we show one typical example which might point to a large class of GUT models. 
We denote the Higgs field which breaks the GUT gauge symmetry as $\sigma$, and the inflaton field $\phi$ is considered a slow-roll background field.
As an analogy, the potential terms relevant for GUT PT take the form, 
\begin{align}
V_{\rm PT}(\phi, \sigma) = D(\phi^2 - \phi_0^2)\,\sigma^2 - \mu \, \sigma^3 + \frac{\lambda}{4}\sigma^4 \,,
\end{align}
where $\phi_0$ is a constant, and $\mu$ and $\lambda$ are either constants or $\phi$-insensitive functions. Here we have not written out the potential for $\phi$ which dominates the inflation and also ignored the influence of $\sigma$ on the evolution of inflation.
During the inflation, the inflaton field $\phi$, as a background, gradually diminishes from a value sufficiently larger than $\phi_0$. The potential along the $\sigma$ direction is minimised at $\sigma = 0$.
Until $\phi$ drops to the critical value $\phi_c \equiv \sqrt{ \phi_0^2 + \frac{\mu^2}{\lambda D}}$, two vacua $\sigma = 0,v_\sigma$ (where $v_\sigma= \frac{3\mu}{2\lambda} + \left[\frac{9 \mu^2}{4\lambda^2} - \frac{2 D}{\lambda} (\phi^2 -\phi_0^2) \right]^{1/2}$) become degenerate. The PT of $\sigma$ from $0$ to $v_\sigma$ should happen at some point for $\phi$ varying from $\phi_c$ to $\phi_0$. 

Gravitational waves are generated during the PT when the energy from the false vacuum to the bulk. The shape of the GW spectrum depends on  gravitational emission modes.
In most studies of PTs in the radiation era, three modes are widely considered: bubble wall collisions, sound waves and magnetohydrodynamic (MHD) turbulence. The bubble wall collisions are collisions of PT fields. Moreover, the expansion of bubbles drives the motion of background fluid (plasma), and the motion modes of plasma in early universe are sound waves and MHD turbulence. They lead the three main sources of GWs during PTs which are widely used in the literature --- bubble wall collisions from PT field, sound waves and MHD turbulence from plasma. For more details on these sources, see recent reviews, e.g., \cite{Caprini:2015zlo, Weir:2017wfa, Bian:2021ini, Athron:2023xlk}.
In our particular case of PT during inflation, since the energy density of plasma is negligible compared with the vacuum energy, GWs induced by sound waves and MHD turbulence are negligible. Therefore, only bubble wall collision is left as the dominant source of GWs via PT during inflation.

We apply the envelope approximation \cite{Kosowsky:1991ua, Kosowsky:1992rz, Kosowsky:1992vn, Kamionkowski:1993fg} to formulate the GW spectrum in the bubble wall collision modes. In this approximation, a fraction $\kappa$ of the latent heat of the PT is deposited in a thin shell close to the PT front. The energy in each shell is assumed to quickly disperse after colliding with another shell such that the energy is primarily stored in the envelope of uncollided shells. 
Numerical simulations in the envelope approximation suggest that the GW spectrum at the moment just after PT follows a broken power law \cite{Huber:2008hg}. Here we briefly summarised the formula below: 
\begin{align}
 \Omega_{\rm GW \star}(f_\star) \equiv \frac{1}{\rho_{\rm tot}} \frac{d \rho_{\rm GW}^{\rm flat}}{d \log f_\star} = \Omega_{\rm GW \star}^{\rm peak}\, \frac{(a+b)(f_\star / f_\star^{\rm peak})^a}{a(f_\star / f_\star^{\rm peak})^{a+b}+b} \,, \label{eq:Omega2}
\end{align}
where the spectral indices $a=2.8$ and $b=1$ are suggested. The broken power law rises as $f^3$ for small frequencies, consistent with general infrared behavior explained by causality \cite{Caprini:2009fx,Cai:2019cdl}, and decreases as $f^{-1}$ for high frequencies. The peak amplitude and peak frequency are respectively given by 
\begin{align}
\Omega_{\rm GW \star}^{\rm peak} &= \frac{0.11v_{\rm w}^3}{0.42+v_{\rm w}^2}\times \kappa^2 \left(\frac{H_\star}{\beta}\right)^2 \left(\frac{\rho_{\rm PT}}{\rho_{\rm tot}}\right)^2 \,, \label{eq:Omega_star_peak}  \\
    f_\star^{\rm peak} &= \frac{0.62\,\beta}{1.8-0.1v_{\rm w}+v_{\rm w}^2} \,, \label{eq:f_star_peak}
\end{align}
where $H_\star^2 = \frac{8}{3}\pi G_{\rm N} \rho_{\rm tot} $ has been used. The peak amplitude is proportional to the square of ${\rho_{\rm PT}}/{\rho_{\rm tot}}$.
$\rho_{\rm PT}$ is the difference of energy density between the true and the false vacua. $\beta^{-1}$ represents the duration of the PT. 
$v_{\rm w}$ is the bubble wall velocity which will be approximated at $v_{\rm w} \approx 1$ in this work. $\kappa$ is the efficiency factor for vacuum energy transformed into kinetic energy of the bulk fluid. Given the ratio of the vacua energy difference  to the radiation energy  $\alpha = \rho_{\rm PT} / \rho_{\rm rad}$, the efficiency factor is calculated to be
\begin{align}
    \kappa(\alpha) &= \frac{0.715\alpha+0.181\sqrt{\alpha}}{1+0.715\alpha} \,. \label{eq:galpha}
\end{align}
This factor approaches to $1$ in the limit $\alpha \to \infty$. 
In the radiation era where $\rho_{\rm tot} = \rho_{\rm PT} + \rho_{\rm rad}$, the peak amplitude is re-written to be
\begin{align}
\Omega_{\rm GW \star}^{\rm peak} &= \frac{0.11v_{\rm w}^3}{0.42+v_{\rm w}^2}\times \kappa^2 \left(\frac{H_\star}{\beta}\right)^2 \left(\frac{\alpha}{\alpha+1}\right)^2 \,,
\end{align}
where is exactly the formula appearing in \cite{Huber:2008hg}. 
However, we keep in mind that there might be additional energy budgets present during the PT. This is exactly what we will discuss soon in the PT during the inflationary era, where the total energy should be $\rho_{\rm tot} = \rho_{\rm Inf} + \rho_{\rm PT} + \rho_{\rm rad}$ and $\rho_{\rm Inf}$ dominates the total energy. 

In our particular case, by assuming $v_{\rm w} \approx 1$ and setting the limit $\alpha \to \infty$, we convert the spectrum $ \Omega_{\rm GW \star}(f_\star) $ to the redshifted one, 
\begin{align}
  &\begin{aligned}
    h^2 \widetilde{\Omega}_{\rm GW}(\tilde{f}) =& 1.27 \times 10^{-6} \times \left(\frac{H_\star}{\beta}\right)^2 \left(\frac{\rho_{\rm PT}}{\rho_{\rm tot}}\right)^2 \left( \frac{100}{g_\star} \right)^{1/3}\\
    &\, \times \frac{(a+b)(\tilde{f} / \tilde{f}^{\rm peak})^a}{a(\tilde{f} / \tilde{f}^{\rm peak})^{a+b}+b} \,, 
  \end{aligned}\nonumber\\
  &\begin{aligned}
    &\tilde{f}^{\rm peak}= 37.8 ~{\rm MHz} \times \left(\frac{\beta}{H_\star}\right) \left( \frac{T_\star}{10^{15}~{\rm GeV}} \right) \left( \frac{g_\star}{100} \right)^{1/6} \,,
  \end{aligned}\label{eq:Omega_tilde_peak}
\end{align}
where $f_\star^{\rm peak}/ \beta \approx 0.23$ at $v_w \approx 1$ has been used. 

\begin{figure}[htbp]
  \centering
  \includegraphics[width=.45\textwidth]{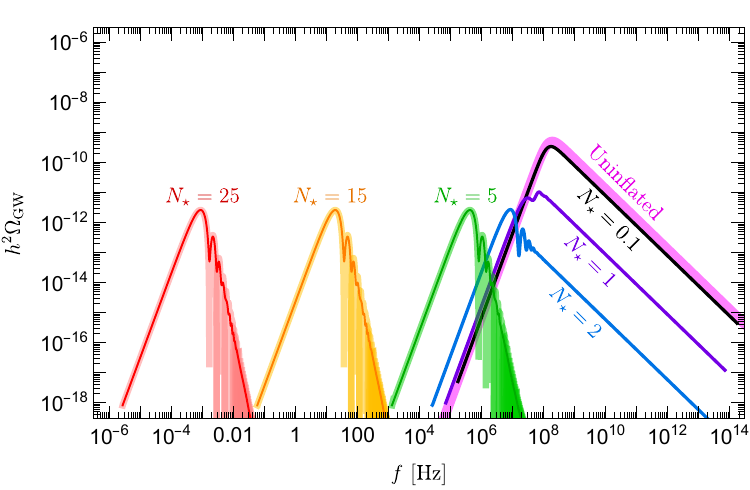}\vspace{3pt}
\caption{Comparison of GW spectra with or without modification by inflation. Inflated GW spectrua for the e-folding number $N_\star =0.1, 1, 2, 5,  15, 25$ are presented. The uninflated GW spectrum is obtained in the envelope approximation with the phase transition temperature assumed to be $T_\star = T_{\rm Rh}$. For $N_\star =5,  15, 25$, curves shown in light and deep colors represent spectra obtained via the instant-source approximation and transitory-source approximation, respectively.   Input parameters: $T_{\rm Rh}=10^{15}$ GeV, $\rho_{\rm PT}/\rho_{\rm tot}=0.1$, $\beta/H_{\star}=5$.} \label{fig:hsqOmega_1}
\end{figure}

Considering the PT as a transitory or instant source at $N_\star$ during the inflation, we include the deformation function $\overline{S}(f)$ or $S(f)$ to the above spectrum, and obtain the inflated GW spectrum $h^2 \Omega_{\rm GW}(f)$
\begin{align} 
   h^2 \Omega_{\rm GW}(f)&=h^2 \widetilde{\Omega}_{\rm GW}(f e^{N_\star}) \times \overline{S}(f) \label{eq:hsqOmega_f}
\end{align} 
following Eq.~\eqref{eq:hsqOmega_master} with the temperature $T_\star$ replaced by the reheating temperature $T_{\rm Rh}$. Here $g_\star$ should be understood as the degree of freedom of radiation in the beginning of the RD era. As a comparison, $h^2 \widetilde{\Omega}_{\rm GW}(\tilde{f})$, with $\tilde{f} = f e^{N_\star}$, is the corresponding GW spectrum not modified by inflation.
For transitory PT, one thinks $1/\beta$ as the approximate duration of sources, so there are $a_\star\Delta_\tau\simeq 1/\beta$ and $\Delta_y\simeq\frac{H_\star}{\beta}\bar{y}$ in Eq.~\eqref{eq:Sbar}.
Actually, for $\beta/H_\star\ge 5$, PT is usually regarded as instant or transitory source of GW, so Eq.~\eqref{eq:hsqOmega_f} is appropriate.
We show in Fig.~\ref{fig:hsqOmega_1} the inflated GW spectra with the e-folding number $N_\star =0.1, 1,2,5, 15, 25$, respectively. The uninflated spectrum, i.e., GWs produced after reheating, is shown as a comparison. For $N_\star \to 0$, inflated GW spectrum can go back to the spectrum without inflation.
As shown, the strength of GWs is suppressed, and its shape is deformed and frequency is redshifted with $N_\star$ increasing from 0 to 5. At $N_\star = 2$, the shape of the spectrum shows clearly distinguishable behaviours in the IR, UV and FUV bands. 
For $N_\star \gtrsim 5$, the GW spectrum in the FUV band is highly suppressed and not shown in the plot. For $N$ varying from 5 to large values, the strength and shape of GW spectrum keep insensitive to the value of $N_\star$, and only the frequency get redshift following the e-folding number $N_\star$. 
The uninflated GWs with peak frequency around 0.1 GHz is shifted to 10 Hz and mHz for $N_\star \simeq 15$ and $25$, respectively. 

\begin{figure}[htbp]
\centering
  \includegraphics[width=.45\textwidth]{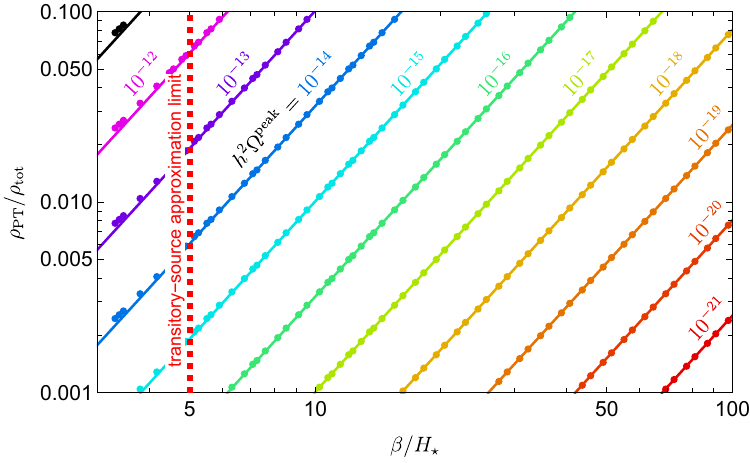}\vspace{3pt}
\caption{This figure is that different peak value of GW spectrum $h^2\Omega_{\rm GW}^{\rm peak}$ is decided by PT velocity $\beta/H_{\star}$ and PT energy proportion $\rho_{\rm PT}/\rho_{\rm tot}$.
Input parameters: $T_{\rm Rh}=10^{15}$ GeV, $N_\star=36$ and $v_{\rm w}=1$.
The points are numerical solutions from Eq.~\eqref{eq:hsqOmega_f}, and the solid lines are approximately analytic by Eq.~\eqref{eq:Omega_peak}.
The region $\beta/H_{\star}\ge 5$ to apply the transitory-source approximation is indicated. 
}\label{fig:Omega_peak}
\end{figure}

\begin{figure}[htbp]
  \centering
  \includegraphics[width=.45\textwidth]{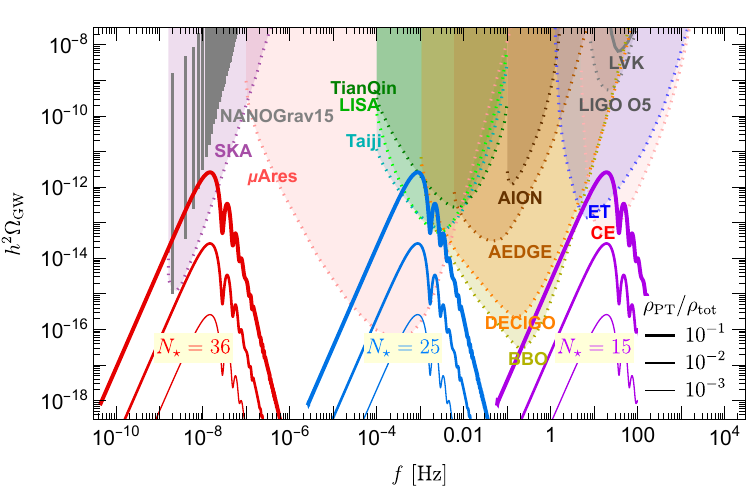}
\caption{Inflated spectra of GWs from GUT phase transition on different e-folding number of time of GW production $N_\star = 15$,  25 and 36 (violet, blue, red, respectively) and different PT energy proportion $\rho_{\rm PT}/\rho_{\rm tot}=0.1,$ 0.01 and 0.001 (curve thickness of thick, middle and thin, respectively).
$\beta/H_{\star}=5$, $T_{\rm Rh}=10^{15}$ GeV and $v_{\rm w}=1$ are taken.}\label{fig:hsqOmega_2}\vspace{3mm}
\end{figure}
\begin{figure}[htbp]
  \centering
  \includegraphics[width=.45\textwidth]{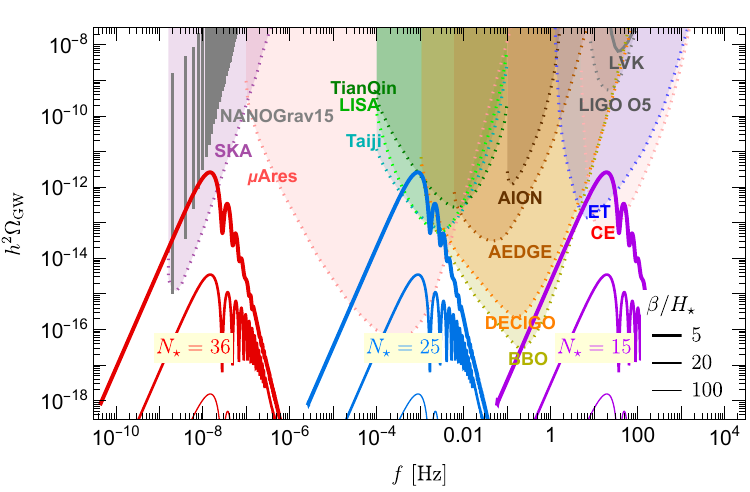}
\caption{Inflated spectra of GWs from GUT phase transition on different e-folds number of time of GW production $N_\star = 15$,  25 and 36 (violet, blue, red, respectively) and inverse duration of PT $\beta/H_{\star} = 5$, 20 and 100 (curve thickness of thick, middle and thin, respectively). $\rho_{\rm PT}/\rho_{\rm tot} = 0.1$, $T_{\rm Rh}=10^{15}$ GeV and $v_{\rm w}=1$ are taken. }\label{fig:hsqOmega_3}
\end{figure}

It is worth mentioning that the peak of the inflated spectrum is not simply a redshift from the peak of the uninflated spectrum. First of all, the peak frequency $f^{\rm peak}$ is almost irrelevant to $\tilde{f}^{\rm peak}$. While the latter, the uninflated peak frequency $\tilde{f}^{\rm peak}$, is proportional to the inverse duration of PT $\beta/H_\star$, the inflated peak frequency $f^{\rm peak}$ is determined by the first peak of the oscillation term in the deformation function ${\cal S}(f)$. For $\beta/H_{\star} >5$, where the transitory-source approximation applies, 
the near peak GW spectrum satisfies 
\begin{align}
h^2\Omega_{\rm GW}(f) \propto f^a S_0(f) \,.
\end{align}
By extremising the RHS of the above formula, we derive $y^{\rm peak} = 2.394$ and $S_0(f^{\rm peak}) = 0.0315$ for $a=2.8$. 
The peak frequency is straightforwardly obtained as
\begin{align}
 f^{\rm peak} \simeq 62.7~\text{MHz} \times  \left(\frac{g_\star}{100}\right)^{1/6}  \left(\frac{T_{\rm Rh}}{10^{15}~{\rm GeV}}\right) \times e^{-N_\star} \,, \label{eq:f_peak}
\end{align}
which is insensitive to the value of $\beta/H_\star$. 
The peak amplitude at this point approximates to
\begin{align}
  \begin{aligned}
    & h^2 \Omega^{\rm peak}_{\rm GW} \simeq S_0(f^{\rm peak})\times h^2 \widetilde{\Omega}_{\rm GW} (f^{\rm peak} e^{N_\star}) \\
    & \simeq 6.27 \times 10^{-7} \times \left(\frac{H_\star}{\beta}\right)^{2+a} \left(\frac{\rho_{\rm PT}}{\rho_{\rm tot}}\right)^2 \left( \frac{100}{g_\star} \right)^{1/3} .
  \end{aligned}\label{eq:Omega_peak}
\end{align}
In Fig.~\ref{fig:Omega_peak}, we show that the analytical approximation matches very well with the numerical calculation. The main point from the analytical formula is the prove of the highly dependence of the peak amplitude on the inverse duration of PT, $\Omega^{\rm peak}_{\rm GW} \propto (\beta/H_\star)^{-2-a} \sim (\beta/H_\star)^{-5}$ and the independence of $N_\star$ (for $N_\star >5$). In order to generate an observable GW signal (e.g., by requiring $\Omega_{\rm GW}^{\rm peak} > 10^{-16}$), $\beta/H_\star \gtrsim 50$ should be roughly required, depending on the fraction $\rho_{\rm PT}/ \rho_{\rm tot}$. 

We further show the inflated spectrum by varying the energy fraction $\rho_{\rm PT}/ \rho_{\rm Inf}$ and the inverse duration of PT $\beta/H$ in Fig.~\ref{fig:hsqOmega_2} and \ref{fig:hsqOmega_3}, respectively. For references, power-law integrated sensitivities for upcoming GW observatories, including space-based laser interferometers (LISA \cite{Audley:2017drz}, Taiji \cite{Guo:2018npi}, TianQin \cite{Luo:2015ght}, BBO \cite{Corbin:2005ny}, DECIGO \cite{Seto:2001qf}, $\mu$Ares \cite{Sesana:2019vho}), atomic interferometers (MAGIS \cite{Graham:2017pmn}, AEDGE \cite{Bertoldi:2019tck}, AION \cite{Badurina:2019hst}), and ground-based interferometers (ET \cite{Sathyaprakash:2012jk}, CE \cite{Evans:2016mbw}), and Square Kilometre Array \cite{Janssen:2014dka} (SKA) are are shown.
$N_\star \simeq 15$ and 25 push the peak frequency to 10 Hz and 1 mHz, respectively. Those are perfect regimes to be tested in ground-based and space-based laser interferometers. The space-based experiments LISA and Taiji will be able to touch the the peak amplitude $10^{-13}$ at the best, corresponding to $\rho_{\rm PT}/\rho_{\rm tot}>0.02$ for $\beta/H_\star=5$ or $\rho_{\rm PT}/\rho_{\rm tot}>0.1$ for $\beta/H_\star = 10$, seen in Fig.~\ref{fig:Omega_peak}. The best sensitivity of GW measurements in the foreseeable future may touch $10^{-16}$, referring to $\rho_{\rm PT}/\rho_{\rm tot}\approx 5\times10^{-4}$ at $\beta/H_\star = 5$ and $\rho_{\rm PT}/\rho_{\rm tot}\approx 0.1$ at $\beta/H_\star = 40$. 
The targeted region based on the NANOGrav 15 data \cite{NANOGrav:2023hvm} and the LIGO/Virgo/KAGRA (LVK) bound \cite{LIGOScientific:2021nrg} set are also presented briefly in the figures. A large e-folding number $N_\star \sim 35$ can shift the frequency to the nHz regime. However, due to the highly suppression of $(\beta/H_\star)^{-2-a}$ and $(\rho_{\rm PT}/ \rho_{\rm tot})^2$, a sufficiently strong signal $h^2 \Omega_{\rm GW}^{\rm peak} \sim 10^{-9}$ is hard to reach for inflated GW via GUT PT in the transitory approximation. However, another mechanism of GW production, i.e., the so-called secondary GWs from curvature perturbation induced by the phase transition during inflation, might provide a signal to matching with the PTA hint \cite{An:2023jxf}.

\subsection{Inflated GWs via phase transition below GUT scale}

\begin{figure}[t!]
      \centering
      \includegraphics[width=.45\textwidth]{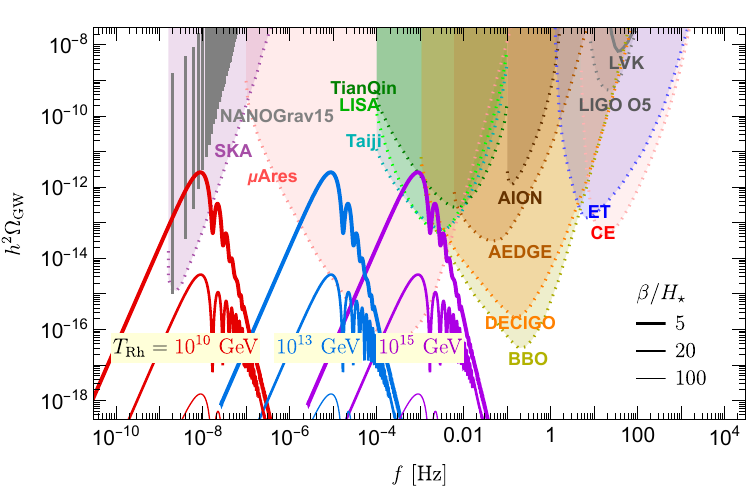}\label{fig:GW_Trh}
  \caption{Inflated spectra of GWs from phase transition at different reheating temperature $T_{\rm Rh}=10^{15}$, $10^{13}$ and $10^{10}$~GeV (shown in violet, blue and red, respectively) and the inverse duration of PT $\beta/H_{\star}=5$, $20$ and $100$ (thick, middle, thin, respectively).
 $N_\star=25$, $\rho_{\rm PT}/\rho_{\rm tot}=0.1$, $v_{\rm w}=1$ are taken as inputs.}
 \end{figure}
\begin{figure}[t!]
  \centering
      \includegraphics[width=.45\textwidth]{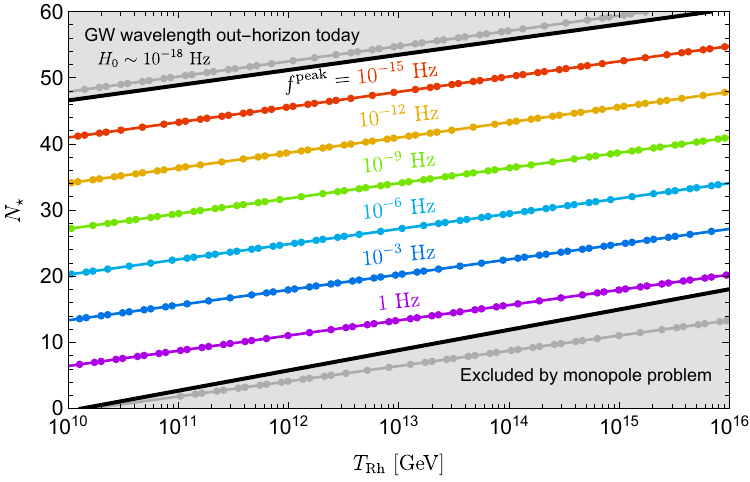}
  \caption{The dependence of the peak frequency $f^{\rm peak}$ of the inflated GW spectrum on the reheating temperature $T_{\rm Rh}$ and the e-folding number $N_\star$ at the phase transition. 
Values of the parameters $\rho_{\rm PT}/\rho_{\rm tot}=0.1$, $\beta/H_{\star}=5$ and $v_{\rm w}=1$ are taken.
The gray region in the bottom-right corner is the excluded region in solving the monopole problem, where the monopole mass is assumed to be at the same scale of the reheating temperature. That in the top-left corner is that the wavelength of GW is outside the horizon today.
  }  \label{fig:fpeak}
\end{figure}

Below the GUT scale, there might be some gauge symmetries persist at some intermediate scales from the GUT breaking to the SM. Some widely studied symmetries, e.g., the Pati-Salam symmetry $SU(4)_c \times SU(2)_L \times SU(2)_R$ via the breaking of $SO(10)$, the 333 model $SU(3)_c \times SU(3)_L \times SU(3)_R$ via the breaking of $E_6$. Their energy scales are naturally assumed to be at a very high scale but not sufficiently as high as the GUT scale. Here we will focus on the scale between $10^{10}$-$10^{15}$~GeV. On the other hand, the breaking of these symmetries generate monopoles with masses around the same scale of the symmetry breaking \cite{Jeannerot:2000sv,Jeannerot:2003qv}, seeing recently in \cite{Chakrabortty:2020otp, Maji:2022jzu}. Monopoles might still a problem if they are heavy enough, and the inflation should be introduced to solve the problem. Below we generalise the discussion in section~\ref{sec:3.2} to inflated GWs via PT below the GUT scale. We relax the reheating temperature $T_{\rm Rh}$ in the range $10^{10}$-$10^{15}$~GeV. The PT for intermediate symmetry breaking is assumed during the inflation with e-folding number $N_\star$ of GUT PT. The monopole mass is again assumed at the same scale of $T_{\rm Rh}$. 

Relaxing the reheating temperature to a lower scale, the inflated GW spectrum formula in Eq.~\eqref{eq:hsqOmega_f} still works. We apply it and show the dependence of the spectrum on the reheating temperature in Fig.~\ref{fig:GW_Trh}. 
It is obvious that $T_{\rm Rh}$ only affects the redshift of frequency and does not change the amplitude and shape of the GW spectrum.
We finally show the dependence of the peak frequency on the reheating temperature as well as the e-folding number $N_\star$ in Fig.~\ref{fig:fpeak}. Both numerical result (dots) based on Eq.~\eqref{eq:hsqOmega_f} and analytical result (lines) in Eq.~\eqref{eq:f_peak} are shown for comparison. 

\section{Conclusions}\label{sec:4}

Phase transition is one main particle source of stochastic gravitational wave (GW) background to be observed in the future GW observatories. 
Phase transition associated with the spontaneous symmetry breaking of Grand unified theories (GUTs) happens at an untouchable high energy scale. The GUT monopole problem requires that the GUT phase transition should happen before or during inflation. In this work, we assume the GUT phase transition happens during inflation. It solves the GUT monopole problem on one hand. On the other hand, GWs from GUT phase transition, if it is first-order, can be redshifted and deformed, and might be observed today in GW observatories. 
Different from some well-known GUT inflationary models where the GUT symmetry breaking is complete at the end of the inflation, we assume the phase transition proceeds very fast at some point during the inflation. The e-folding number of the phase transition is denoted as $N_\star$. We study the general behaviour of the inflated GW spectrum and its application in GUT phase transition during inflation. 

We derived the general formalism of GW spectrum generated from any instant or transitory source during inflation. The general correlation between the inflated and uninflated (i.e., GWs generated after inflation) GW spectra is given in Eq.~\eqref{eq:hsqOmega_master}. Beside the redshift of frequency, another important modification of the GW spectrum is the deformation. It changes the shape of spectrum. We denote the deformation function as $S(f)$ with $f$ the frequency today.  We give a complete analytical description of the deformation function for the e-folding number $N_\star$ taking any reasonable value $0< N_\star \lesssim 60$. For $N_\star >5$, we recover the result obtained in \cite{An:2020fff}. 
We clarified IR and UV regimes by comparing the frequency nowadays $f$ with $\frac{a_\star H_{\star}}{2\pi a_0}$, and $f \gg \frac{a_{\rm Rh} H_{\star}}{2\pi a_0}$ is further regarded as the FUV regime. Properties of the deformation function in these regimes were discussed. Here, $a_\star$, $a_{\rm Rh}$, and $a_0$ are co-moving factors at instant or transitory source, at reheating, and today, respectively, and $H_\star$ is the Hubble rate at instant or transitory-duration source, which is identical to the Hubble rate during inflation. In the IR band, the deformation is frequency-independent, providing only an overall suppression factor $1/9$ for $N_\star > 5$. In the UV band, $S(f)$ oscillates along $f$. In the FUV band, $S(f)$ provides an suppression inversely proportional to the quartic square of the co-moving factor. 

We apply the formalism to GUT phase transition during inflation. The transitory-source approximation holds with the assumption  $\beta/H_\star >5$, where $\beta$ is the inverse duration of phase transition. The inflated GW spectrum, after deformed by inflation, shows the peak of the spectrum at $f^{\rm peak} \simeq 60~\text{MHz} \cdot (\frac{T_{\rm Rh}}{10^{15}~{\rm GeV}})  e^{-N_\star}$ and the peak amplitude highly suppressed by a factor $\sim (\beta/H_\star)^{-5}$. This peak is not simply a redshift of the peak of uninflated GW spectrum obtained by phase transition in the radiation era. For $N_\star \simeq 15$ and 25, inflation pushes the peak frequency of GWs from 0.1 GHz to 10Hz and mHz hands, respectively, referring to the favoured regimes for ground-based and space-based laser interferometers, respectively. LISA/Taiji will be able to test GUT phase transition with $\rho_{\rm PT}/\rho_{\rm tot} \sim {\cal O}( 0.1)$ and $\beta/H_\star \lesssim 10$. The best sensitivity of GW measurements in the foreseeable future may be able to test $\rho_{\rm PT}/\rho_{\rm tot} \approx {\cal O}(10^{-4})$ or $\beta/H_\star \lesssim 40$.  A larger e-folding number $N_\star \simeq 36$ can further push the GWs to the nHz band, but the suppressed amplitude cannot explain the hint of PTA data. 
We further generalised the discussion to inflated GWs via phase transition below the GUT scale by relaxing the reheating temperature. 
In summary, the inflated GW spectrum shows a clear correlation between the peak frequency and the e-folding number. If any of these experiments measures a inflated spectrum at a certain peak frequency, we will be able to track the e-folding number between GUT phase transition and the end of inflation. On the other hand, predicting a special value of this e-fold number deserves further exploration in an explicit inflationary model building and will be carried out in our future study.

\acknowledgments

Y.L.Z. would like to thank useful discussions with H.P. An. This work was supported by the National Natural Science Foundation of China (NSFC) under Grants No.~12205064, No.~12347103, and Zhejiang Provincial Natural Science Foundation of China under Grant No. LDQ24A050002.

\bibliographystyle{apsrev4-2}
\bibliography{apsprd}

\end{document}